# On the relationships between bibliographic characteristics of scientific documents and citation and Mendeley readership counts: A large-scale analysis of Web of Science publications[1]

Zohreh Zahedi* & Stefanie Haustein**

* z.zahedi.2@cwts.leidenuniv.nl
*Centre for Science and Technology Studies (CWTS), Leiden University, Wassenaarseweg 62A, Leiden, 2333 AL (The Netherlands)

**stefanie.haustein@umontreal.ca
**École de bibliothéconomie et des sciences de l'information, Université de Montréal, C.P. 6128, Succ. Centre-Ville, Montréal, H3C 3J7 (Canada)

**Abstract**

In this paper we present a first large-scale analysis of the relationship between Mendeley readership and citation counts with particular documents' bibliographic characteristics. A data set of 1.3 million publications from different fields published in journals covered by the Web of Science (WoS) has been analyzed. This work reveals that document types that are often excluded from citation analysis due to their lower citation values, like editorial materials, letters, or news items, are strongly covered and saved in Mendeley, suggesting that Mendeley readership can reliably inform the analysis of these document types. Findings show that collaborative papers are frequently saved in Mendeley, which is similar to what is observed for citations. The relationship between readership and the length of titles and number of pages, however, is weaker than for the same relationship observed for citations. The analysis of different disciplines also points to different patterns in the relationship between several document characteristics, readership, and citation counts. Overall, results highlight that although disciplinary differences exist, readership counts are related to similar bibliographic characteristics as those related to citation counts, reinforcing the idea that Mendeley readership and citations capture a similar concept of impact, although they cannot be considered as equivalent indicators.

**Keywords**
Mendeley readership; WoS Citation; Bibliographic characteristics; Document types

---

**Introduction**

*Effect of document characteristics on citation impact*

Measuring research impact using citation analysis has a long tradition in the field of scientometrics. Today, citation-based indicators are widely used and play a central role in the evaluation of scientific works. Despite their *de facto* use as proxies of scientific quality, citations are not able to fully capture the use and influence of scientific papers (Moed, 2005; MacRoberts & MacRoberts, 2017). Bibliometric research has also shown that a variety of factors can influence citation counts (Opthof & Leydesdorff, 2010; Waltman et al., 2011; Larivière & Gingras, 2011). Such factors include, the document types and age of publications, their number of pages, the length of their titles and reference lists (Bornmann & Leydesdorff, 2015; Bornmann, Leydesdorff, & Wang, 2014; Vieira & Gomes, 2010); their different theoretical or methodological approaches (Antonakis, et al., 2014); whether they are open access (Hajjem, Harnad, & Gingras, 2006); the citation propensity of their fields and their interdisciplinarity (Yegros-Yegros, Rafols, & D'Este, 2015); or the Impact Factor of their publication journal (Boyack & Klavans, 2005).

Numerous previous studies have analyzed whether citation impact is affected by various document characteristics. These studies have explored different characteristics at the article, journal, and author levels using correlation and regression analyses. For example, in the Natural, Life, and Health sciences (Thelwall, 2017), papers with unusual and obscure titles were associated with lower citation impact. Mixed results were found regarding the effect of title length (Stremersch, et al., 2015; Jacques & Sebire, 2010), or titles that included non-alphanumeric characters such as hyphens or colons (Buter & Van Raan, 2011; Haslam, et al., 2008; Nair & Gibbert, 2016). Based on the assumption that longer articles with longer reference lists may reflect in-depth analysis and diversity of ideas, the number of pages and references have also been analyzed as factors that may affect citation counts (Fox & Boris, 2016). The results showed that papers with more references and more pages tended to get more citations (Ajiferuke & Famoye, 2015; Davis, et al., 2001). Similarly, the number of authors, institutes, and countries involved in a given publication may indicate the extent of collaboration, which is again assumed to increase citation impact. However, results regarding the effect of collaboration on citation rates are mixed (for an overview see Onodera & Yoshikane, 2015) as regards variations by country of collaboration (Thelwall & Sud, 2016), level of collaboration (e.g. whether national, international, intra/inter institutional) (Leimu & Koricheva, 2005), or authors and disciplines (Williams, et al., 2009). For a recent review of studies analyzing factors affecting citation counts we refer to Tahamtan, Safipour Afshar, & Ahamdzadeh (2016).

*Effect of document characteristics on social media visibility*

In the context of recently introduced altmetrics—or, more specifically, its subset of social media based metrics (e.g., Facebook, Twitter, blogs, Wikipedia, Mendeley)—the effect to which some factors influence social media activity remains understudied. One large-scale study examining the effect on social media metrics of typical document characteristics (including document type, discipline, number of pages, title length, number of references, and collaboration patterns) conducted by Haustein, Costas, & Larivière (2015). This study was based on Altmetric.com and Web of Science data and found that although effects were weaker than for citations, documents were more likely to be tweeted if they had longer reference lists and involving a greater number of authors, institutes, and countries. Correlations between social media metrics and document characteristics were, however, quite low to non-existent, which was mostly due to the skewed nature of social media events related to journal articles, with most of them having no metrics at all. Social media metrics (particularly Facebook and Twitter counts) correlated mostly among each other, indicating a circular relationship (Bourdieu, 1998), meaning that being picked up by one social media increases the chances of being picked up by another one. Haustein et al. (2015) also found that news items and editorials were among the most tweeted document types, which indicates that outputs that contain more condensed, novel, opinion-based and easy-to-understand pieces tend to be more popular on Twitter. The results contrast with the citation patterns for these types of documents, which are substantially less cited than articles and reviews. Overall, the study by Haustein et al. (2015) showed that characteristics that typically are related to higher citation counts had a smaller relationship with social media counts, sometimes even in an entirely different manner (for instance, longer titles were associated with higher citation counts but with lower Twitter mentions).

*Mendeley readership and citation counts*

Mendeley is an online reference manager that allows users to save documents in their own libraries and share their libraries with others. Statistics about how often a particular document is saved are made available via the Mendeley API as 'readership' counts. While this count is described by Mendeley as 'readership', it does not actually indicate that the user who saved the document has actually 'read' it, but simply that the user has saved the reference in the library. As such, Mendeley 'saves' are seen more as acts of access to documents than of their appraisal (Haustein, Bowman & Costas, 2016), indicating that the level of engagement captured by these acts is very low.

However, Mendeley has been identified as the most prevalent and noteworthy altmetric source. It has been found that readership counts often exceed citations, and that there is a high representation of recent publications on the platform (Thelwall & Sud, 2015). Compared to other altmetric indicators, Mendeley readership counts were shown to have moderate to strong correlations with citation counts (for a review see Sugimoto, et al., 2016), which reflects a greater similarity with citations than other altmetric indicators (Costas, Zahedi, & Wouters, 2015a). This can be explained by the large numbers of academic users in Mendeley, and the frequent use of Mendeley in a pre-citation context (Mohammadi, Thelwall, & Kousha, 2016). The number of Mendeley users who have added an article to their libraries has been suggested as an early indicator of citation impact (Thelwall & Sud, 2015), and Mendeley itself has been identified as a relevant tool with which to identify highly cited publications (Zahedi, Costas, &Wouters, 2017). Mendeley readership distributions have also been shown to be very similar to citation distributions (Costas, Haustein, Zahedi, & Larivière, 2016), and it has been suggested that field-normalized readership scores could be calculated in a similar fashion as for citations (Bornmann & Haunschild, 2016).

Given these similarities between readership and citations, one might expect that Mendeley readership counts are also related to the same document characteristics as citations. While some characteristics, such as document age (Thelwall, Haustein, Larivière, & Sugimoto, 2013), disciplines (Haustein et al., 2015), topics (Costas, Zahedi, & Wouters, 2015a), or countries (Alperin, 2015) have been already explored for Mendeley, a systematic study of other quantitative document characteristics previously investigated for citations is still lacking in the literature regarding Mendeley readership. A recent study (Didegah, Bowman, & Holmberg, in press[2]) investigated the relationship between some factors (such as JIF, cited references, title length, country and institute's prestige, and field type and size) with Mendeley and citation counts for a sample of Finnish WoS papers. However, a global large-scale disciplinary analysis of the relationship of these characteristics as well as some other factors is still missing in the literature.

The present work represents the first large-scale analysis of the relationship between Mendeley readership and specific documents' bibliographic characteristics. Specifically, this study aims to improve the understanding of the relationship between Mendeley readership and selected document characteristics, including document types, number of pages, title length, length of reference list, and number of authors, institutes, and countries of the papers. We study how the relationship between these bibliographic characteristics and readership is similar to and/or differs from that observed for citations and, how this relation varies across different fields.

The selection of document characteristics represents only a limited number of quantitative variables, and we acknowledge that it does not consider other qualitative aspects that might affect the extent to which articles attract users on Mendeley. Nevertheless, this study will contribute to a better understanding of how Mendeley readership relates to basic document characteristics by providing a clear framework of this relation. This could contribute to the identification of document-related differences between Mendeley readership and citations that can help the future construction of appropriate and meaningful indicators based on Mendeley readership. The study builds on the work by Haustein et al. (2015) and Zahedi et al. (2016) by studying the same document characteristics, while taking into account longer citation and readership windows. Also, this study improves on these previous studies by including Mendeley readership counts and documents from several different disciplines and by using more advanced regression analysis (in contrast with the more basic correlation analysis employed in previous studies). The paper addresses the following research questions:

- To what extent do document type and document characteristics (i.e., number of pages, length of title and reference list, and number of authors, institutes and countries involved) associate with the number of Mendeley readership?
- How do these relationships between document characteristics and Mendeley readership vary across disciplines? And how do they compare with those observed for citations?

**Data and Methodology**

This study compares Mendeley readership to citation counts received by 2012 Web of Science (WoS) publications with a Digital Object Identifier (DOI) from all disciplines (n=1,339,279). Citation counts from the Centre for Science and Technology Studies (CWTS) in-house WoS database were collected through the end of August 2015. Mendeley readership counts were extracted from the Java Script Object Notation (JSON) files obtained from querying the Mendeley Application Programming Interface (API) using DOIs in July 2015. The analyzed document properties included document type (as recorded by WoS), number of pages, number of cited sources in the reference list (including non-source items), number of characters in the title, number of authors, institutes, and countries of the paper, as well as the scientific disciplines (according to the Leiden Ranking (LR) classification based on the CWTS in-house version of WoS).

---

[2] This study uploaded to arXiv (https://arxiv.org/abs/1710.08594) while our paper was under review by the Journal of Informetrics.

General descriptive statistics were computed for all documents (n= 1,339,279). This included the percentage of papers with at least one citation or one Mendeley readership count (coverage) and the average number of counts per paper (density). To assess the influence of each of the independent variables (i.e., title length, number of pages, number of references, authors, institutes and countries) on readership and citation counts, a standard linear regression (Ordinary Least Squares) analysis was performed using RStudio for all articles and reviews (n= 1,197,162). The OLS analyses were performed to detect if there were any disciplinary differences in the factors influencing readership vs. citation impact[3]. Although there are some debates surrounding how to choose a proper regression model for skewed distributions (see Ajiferuke & Famoye, 2015), the OLS analysis (after log-transforming readership and citation counts[4]) has been shown to be the most suitable regression strategy for citation data and altmetrics (Thelwall & Wilson, 2014). This method proved to be more reliable than negative binomial (NB) or zero inflated negative binomial (ZINB) regression analyses (Thelwall & Wilson, 2014) as it takes into account very high values which are typical for skewed distributions (Thelwall, 2016), which is similar to the skewed log normal distributions of both citation and readership counts in this study.

**Results**

*Distribution of citation and readership across different disciplines*
Out of the 1,339,279 WoS publications, 93.6% (n= 1,254,852) were found in Mendeley via the DOI (Table 1). Of the total publications, 81.7% (n= 1,094,166) were cited at least once in WoS and 84.2% (n= 1,127,849) had at least one readership (at time of data collection). This set of publications accumulated 14,732,103 readership counts and 10,289,891 citation counts in total. Table 1 presents the most important results regarding the coverage and density of both counts in our database. On average, each paper received approximately 7.7 citations and 11 readers, which indicates that documents have been saved more in Mendeley than cited. Regarding the share of publications with at least a Mendeley readership and a citation, they are very close in most disciplines with the exception of the Social sciences and humanities, in which 81.7% of publications had been saved on Mendeley, while only 64.1% had been cited (Table 1).

On average, papers from the Social sciences and humanities had 14.1 readership counts on Mendeley and were cited only 4 times in WoS. The Social sciences and humanities represent the discipline with the second highest readership density after Life and earth sciences (16.5 mean readership vs. 8.6 mean citation counts) and Biomedical and health sciences (12.2 mean readership vs. 8.7 mean citation counts). In Natural sciences and engineering (8.8 mean readership vs. 9 mean citation counts) and Mathematics and computer science (7.8 mean readership vs. 4.4 mean citation counts) Mendeley readership counts are lower. Natural sciences and engineering is the only discipline where average citation counts (9) slightly exceed average Mendeley readership counts (8.8) (Table 1).

---

[3] Only total readership counts (not the readership disaggregated by Mendeley users) are included in the regression model.
[4] The value of 1 was added to readership and citation counts to include papers without zero values in the regression.

**Table 1.** Mendeley and citation coverage and density (standard deviation) per discipline.

| | | All disciplines<br>N=1,339,279 | Biomedical & health sciences<br>N=595,254 | Life & earth sciences<br>N=254,817 | Mathematics & computer science<br>N=135,445 | Natural sciences & engineering<br>N=413,862 | Social sciences & humanities<br>N=159,389 |
|---|---|---|---|---|---|---|---|
| **Citations** | Coverage | 81.7 | 83.4 | 89.1 | 75.4 | 87.3 | 64.1 |
| | Density | 7.7 | 8.7 | 8.6 | 4.4 | 9.0 | 4.0 |
| | *Std. dev.* | *18.8* | *20.9* | *19.0* | *12.4* | *21.3* | *8.7* |
| **Mendeley Readership** | Coverage | 84.2 | 86.5 | 91.4 | 76.4 | 83.7 | 81.7 |
| | Density | 11.0 | 12.2 | 16.5 | 7.8 | 8.8 | 14.1 |
| | *Std. dev.* | *23.9* | *26.5* | *32.5* | *29.5* | *18.0* | *27.0* |

*Analysis of relation between document types and their citation and readership impact*

Figure 1 presents the coverage and average values of both citation and readership for the most important document types in WoS. Reviews and articles are the document types that were most commonly cited and saved on Mendeley. This is to be expected since these types of publications represent the most important research findings that are widely relevant to various audiences. Although both coverage and density were higher for reviews and articles, editorial materials (6.5 density; 68.2% coverage), and news items (5 density; 59.4% coverage) were also frequently saved by Mendeley users, followed by letters (2.7 density; 58.4% coverage), book reviews (1.5 density; 28.8% coverage), biographical items (0.8 density; 28.4% coverage), corrections (3.4 density; 25.4% coverage), and meeting abstracts (0.6 density; 24.7% coverage).

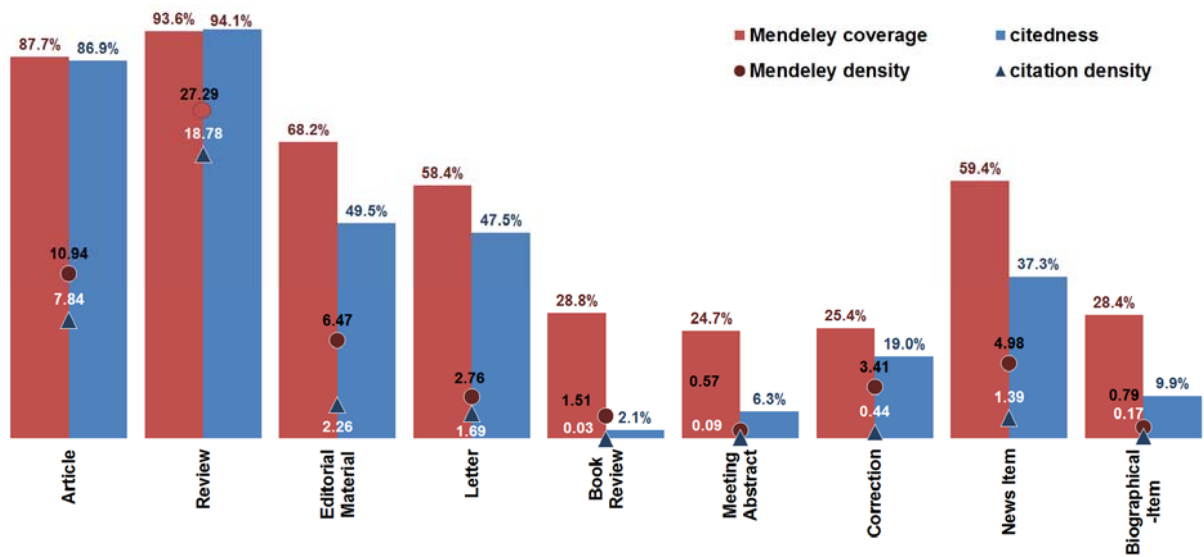

**Figure1.** Mendeley coverage and density, citedness and citation density per document type. Document types appear in order of frequency: article (N=1,132,428), review (N=64,734), editorial material (N=60,533), letter (N=29,410), book review (N=21,710), meeting abstract (N=13,071), correction (N=9,817), news item (N=4,880) and biographical item (N=2,302). Document types that occurred less than 2,300 times were not analyzed.

*Analysis of relations between document characteristics and citation and readership impact across disciplines*

General average values and standard deviation of the examined variables are presented in Table A1 in the appendix. This table illustrates that, on average, papers from the Life and earth sciences, Biomedical and health sciences, and Natural sciences and engineering tend to have the longest titles. Also, these papers are more collaborative than those from both Social sciences and humanities and Mathematics and computer science in terms of number of authors involved. Papers from Social sciences and humanities have the longest reference lists and number of pages.

The results of the OLS regression analysis among the dependent and independent variables of articles and reviews (n=1,197,162) and their individual disciplines[5] are presented in Table 2. Multicollinearity was tested before making inferences about the coefficient estimates using Variance Inflation Factor (VIF). VIF is an indicator of how much the variance of estimated coefficients change due to multicollinearity, although there is no consensus of what is a high VIF value. VIF value above 10 is often regarded as an indicator of serious multicollinearity (Tabachnick & Fidell, 2007) and a VIF value above 4 is typically considered problematic (Rovai, Baker, & Ponton, 2013). As a result, the number of authors and institutes from Natural sciences and engineering and their interpretations have been excluded from the OLS model due to their high VIF values (>10). Number of institutes in all other fields have VIF values between 2.1 to 2.7[6], while all other variables have VIF values below 1.9; hence no collinearity is expected.

**Table 2.** Ordinary Least Squares regression analysis (OLS) between dependent (citation and readership) and independent (title length, number of references, pages, authors, institutes, and countries) variables across all articles and reviews from their individual LR fields.

### a) Dependent variable Log (Citation +1)

**Biomedical & health sciences (n= 502,610)**

| $R^2$=0.152 | Coefficients (B) | B | Std. Error | CI 95% (B) | | VIF |
|---|---|---|---|---|---|---|
| Number of references | 0.010*** | 0.311 | 0.000 | 0.009 | 0.010 | 1.309 |
| Title length | 0.001*** | 0.046 | 0.000 | 0.001 | 0.001 | 1.036 |
| Number of pages | 0.004*** | 0.027 | 0.000 | 0.004 | 0.004 | 1.295 |
| Number of authors | 0.033*** | 0.169 | 0.000 | 0.033 | 0.034 | 1.907 |
| Number of institutes | -0.009*** | -0.022 | 0.001 | -0.011 | -0.007 | 2.607 |
| Number of countries | 0.111*** | 0.097 | 0.002 | 0.108 | 0.115 | 1.670 |

**Life & earth sciences (n=242,117)**

| $R^2$=0.140 | Coefficients (B) | B | Std. Error | CI 95% (B) | | VIF |
|---|---|---|---|---|---|---|
| Number of references | 0.011*** | 0.347 | 0.000 | 0.011 | 0.011 | 1.453 |
| Title length | 0.000*** | -0.014 | 0.000 | 0.000 | 0.000 | 1.008 |
| Number of pages | -0.014*** | -0.094 | 0.000 | -0.015 | -0.013 | 1.444 |
| Number of authors | 0.037*** | 0.163 | 0.001 | 0.036 | 0.038 | 1.678 |
| Number of institutes | -0.001† | -0.002 | 0.002 | -0.004 | 0.002 | 2.651 |
| Number of countries | 0.090*** | 0.078 | 0.003 | 0.084 | 0.096 | 1.864 |

---

[5] Bootstrap and 95% confidence interval showed significant results although the power (R-squared) is not very strong which is common in social sciences studies. This means that although variables in the model have statically significant coefficients, they account for only a small portion of the variation in each dependent variables separately.

[6] Although the spearman correlations among the number of institutes and authors and the number of authors and countries are not very high ranging from .4 to .6 across fields (see Table A2 in the appendix), the results for number of institutes should be interpreted with caution (due to its higher VIF value than other variables).

**Natural sciences & engineering (n=404,457)**

| $R^2$=0.148 | Coefficients (B) | B | Std. Error | CI 95% (B) | | VIF |
|---|---|---|---|---|---|---|
| Number of references | 0.014*** | 0.401 | 0.000 | 0.014 | 0.014 | 1.274 |
| Title length | 0.002*** | 0.051 | 0.000 | 0.001 | 0.002 | 1.005 |
| Number of pages | -0.017*** | -0.106 | 0.000 | -0.017 | -0.016 | 1.288 |
| Number of countries | 0.076*** | 0.091 | 0.001 | 0.073 | 0.078 | 1.015 |

**Mathematics & computer science (n= 131,220)**

| $R^2$=0.137 | Coefficients (B) | β | Std. Error | CI 95% (B) | | VIF |
|---|---|---|---|---|---|---|
| Number of references | 0.015*** | 0.312 | 0.000 | 0.015 | 0.016 | 1.148 |
| Title length | 0.002*** | 0.062 | 0.000 | 0.002 | 0.002 | 1.040 |
| Number of pages | -0.008*** | -0.073 | 0.000 | -0.009 | -0.008 | 1.162 |
| Number of authors | 0.038*** | 0.115 | 0.001 | 0.036 | 0.040 | 1.464 |
| Number of institutes | 0.006* | 0.008 | 0.003 | 0.000 | 0.012 | 2.195 |
| Number of countries | 0.096*** | 0.062 | 0.005 | 0.086 | 0.106 | 1.703 |

**Social sciences & humanities (n=125,795)**

| $R^2$=0.195 | Coefficients (B) | β | Std. Error | CI 95% (B) | | VIF |
|---|---|---|---|---|---|---|
| Number of references | 0.010*** | 0.312 | 0.000 | 0.010 | 0.010 | 1.224 |
| Title length | 0.001*** | 0.047 | 0.000 | 0.001 | 0.002 | 1.072 |
| Number of pages | -0.018*** | -0.150 | 0.000 | -0.018 | -0.017 | 1.265 |
| Number of authors | 0.092*** | 0.232 | 0.001 | 0.089 | 0.094 | 2.136 |
| Number of institutes | 0.026*** | 0.037 | 0.003 | 0.020 | 0.032 | 2.723 |
| Number of countries | 0.064*** | 0.043 | 0.005 | 0.055 | 0.074 | 1.644 |

**b) Dependent variable Log (Readership +1)**

**Biomedical & health sciences (n= 502,610)**

| $R^2$=0.122 | Coefficients (B) | β | Std. Error | CI 95% (B) | | VIF |
|---|---|---|---|---|---|---|
| Model | | | | | | |
| Number of references | 0.010*** | 0.302 | 0.000 | 0.010 | 0.010 | 1.309 |
| Title length | -0.001*** | -0.023 | 0.000 | -0.001 | -0.001 | 1.036 |
| Number of pages | 0.004*** | 0.026 | 0.000 | 0.004 | 0.005 | 1.295 |
| Number of authors | 0.005*** | 0.022 | 0.000 | 0.004 | 0.005 | 1.907 |
| Number of institutes | 0.010*** | 0.001 | 0.002 | 0.008 | 0.011 | 2.607 |
| Number of countries | 0.131*** | 0.106 | 0.002 | 0.127 | 0.135 | 1.670 |

**Life & earth sciences (n=242,117)**

| $R^2$=0.140 | Coefficients (B) | β | Std. Error | CI 95% (B) | | VIF |
|---|---|---|---|---|---|---|
| Number of references | 0.013*** | 0.367 | 0.000 | 0.013 | 0.013 | 1.453 |
| Title length | -0.002*** | -0.076 | 0.000 | -0.003 | -0.002 | 1.008 |
| Number of pages | -0.019*** | -0.114 | 0.000 | -0.020 | -0.019 | 1.444 |
| Number of authors | 0.012*** | 0.044 | 0.001 | 0.010 | 0.013 | 1.678 |

| | Coefficients (B) | β | Std. Error | CI 95% (B) | | VIF |
|---|---|---|---|---|---|---|
| Number of institutes | 0.025*** | 0.040 | 0.002 | 0.021 | 0.029 | 2.651 |
| Number of countries | 0.132*** | 0.098 | 0.003 | 0.125 | 0.139 | 1.864 |

**Natural sciences & engineering (n=404,457)**

| $R^2$=0.092 | Coefficients (B) | β | Std. Error | CI 95% (B) | | VIF |
|---|---|---|---|---|---|---|
| Number of references | 0.012*** | 0.329 | 0.000 | 0.012 | 0.012 | 1.274 |
| Title length | -0.001*** | -0.025 | 0.000 | -0.001 | -0.001 | 1.005 |
| Number of pages | -0.015*** | -0.088 | 0.000 | -0.015 | -0.014 | 1.288 |
| Number of countries | 0.035*** | 0.040 | 0.001 | 0.032 | 0.038 | 1.015 |

**Mathematics & computer science (n= 131,220)**

| $R^2$=0.216 | Coefficients (B) | β | Std. Error | CI 95% (B) | | VIF |
|---|---|---|---|---|---|---|
| Number of references | 0.026*** | 0.447 | 0.000 | 0.026 | 0.027 | 1.148 |
| Title length | 0.000† | 0.001 | 0.000 | 0.000 | 0.000 | 1.040 |
| Number of pages | -0.018*** | -0.132 | 0.000 | -0.019 | -0.017 | 1.162 |
| Number of authors | 0.048*** | 0.118 | 0.001 | 0.045 | 0.050 | 1.464 |
| Number of institutes | 0.017*** | 0.017 | 0.004 | 0.010 | 0.024 | 2.195 |
| Number of countries | 0.057*** | 0.031 | 0.006 | 0.045 | 0.069 | 1.703 |

**Social sciences & humanities (n=125,795)**

| $R^2$=0.169 | Coefficients (B) | β | Std. Error | CI 95% (B) | | VIF |
|---|---|---|---|---|---|---|
| Number of references | 0.013*** | 0.342 | 0.000 | 0.013 | 0.013 | 1.224 |
| Title length | 0.002*** | 0.049 | 0.000 | 0.002 | 0.002 | 1.072 |
| Number of pages | -0.024*** | -0.175 | 0.000 | -0.025 | -0.023 | 1.265 |
| Number of authors | 0.067*** | 0.144 | 0.002 | 0.063 | 0.070 | 2.136 |
| Number of institutes | 0.017*** | 0.020 | 0.004 | 0.010 | 0.024 | 2.723 |
| Number of countries | 0.128*** | 0.072 | 0.006 | 0.117 | 0.139 | 1.644 |

Statistical significance: *p < .05; **p < .01; ***p < .001, † shows not significant coefficient (p>0.05); $R^2$ = amount of variance explained by IVs, B = Unstandardized coefficient, β = Standardized coefficient (values for each variable are converted to the same scale so they can be compared), SE = Standard Error, CI = Confidence Interval, VIF= Variance Inflation Factor.

Based on Table 2, number of cited references, authors, institutes, and countries associate with increased citation and readership counts across all fields. There are two exceptions: first, the relation between number of institutes is not significant (sig=0.55, p>0.05) in Life and earth sciences; second, in Biomedical and health sciences, number of institutes is not associated with high citation impact since the effect is very low[7] (Table 2). Longer papers associate with increased citation and readership counts in Biomedical and health sciences, while they associate with decreased citation and readership counts in all other fields. In terms of title length, papers with longer titles are cited more than those with shorter titles. This is supported by both the OLS analysis and the correlation analysis (see Table A2 in the appendix), since citations are positively related to the number of characters of papers' titles. Yet, it seems that this is not the case for Mendeley readership, since OLS and correlations tend to be negative. The only exception is seen for Social sciences and humanities, where papers with longer titles are saved more than those with shorter tiles in Mendeley. The relationship between title length and readership impact in Mathematics and computer science is not significant (sig=0.65, p>0.05).

---

[7] The results should be interpreted with caution due to its higher VIF value (VIF=2.6) than other variables.

**Discussion and conclusions**

In this paper, we present the results of a first large-scale analysis of the relationship between document characteristics and Mendeley readership and citation counts for a large set of WoS publications. The aim of this analysis was to test how certain document characteristics are related to citation and readership counts, and how these patterns differ across fields.

The results showed that papers are overall saved in Mendeley more than cited. Activity on social bookmarking platforms such as Mendeley is expected to exceed citations, given the shorter time lag and broader type of use of saving (reading) over citing—not all that is read is cited—as well as the conceptually larger audience of readers as compared to citing authors. Also, the process of saving documents in Mendeley involves less time and engagement than reading, understanding, and citing them in a paper. The higher density of readership than citation was particularly evident in Social sciences and humanities. This reflects how the citation culture of this discipline is strongly influenced by the coverage of WoS, which excludes citations from books and regional and non-English language journals, that play central role in the Social sciences and humanities. Moreover, the citation delay is particularly high in the Social sciences and humanities, so that only a fraction of citations will have appeared during the citation window of 40 months captured in this study. In contrast, Mathematics and computer science is the field with the lowest Mendeley readership density. The same low presence of Mendeley readers in Mathematics compared to other fields has also been reported in the study by Thelwall (in press). This might suggest that other reference managers such as JabRef, EndNote, or BibSonomy could be more popular in this field or that scholars from this field do not use any reference managers at all. Studies have found that in general uptake of Mendeley is low among researchers (Bowman, 2015; Mas-Bleda, Thelwall, Kousha, & Aguillo, 2014; Van Noorden, 2014). The study by Haustein et al. (2015) also showed lower social media mentions from Twitter, Facebook or blogs for papers from Mathematics and computer science than those from Social sciences and humanities, Biomedical and health sciences, and Life and earth sciences. This supports the idea that social, health, and environmental topics receive more attention in social media than highly technical topics (Costas, Zahedi, & Wouters, 2015b).

Our study shows that reviews and articles are the most prevalent document types in Mendeley. This is to be expected, since these document types represent the most important research findings that are widely relevant to various audiences. However, compared with citations, Mendeley readership exhibits higher coverage and relatively higher density values for other document types (e.g., editorial materials, letters, and news items). This indicates that although they do not contain material that is frequently cited, these document types are informative and do attract Mendeley users. This supports the idea that Mendeley readership could be useful for the analysis and evaluation of these document types, especially given that most of them are traditionally excluded from citation analysis (Waltman et al., 2011) due to their lower citation frequency. The popularity of these other document types has been reported for other altmetrics, particularly for Twitter (Haustein et al., 2015) and research blogs (Schema, Bar-Ilan, & Thelwall, 2015). This stronger presence of other document types in Mendeley also suggests that Mendeley readership may capture a mix of citation and social media patterns, since documents with different relevance for citations and social media are reasonably saved in Mendeley (e.g. news media items or editorial material).

Regarding the relationship between various document characteristics and Mendeley readership and citation counts, similar patterns were found in most cases. Akin to citations, papers with more references, written by many authors from several institutes and countries, are saved more frequently in Mendeley. This is because collaborative papers based on a large number of cited references may reflect more in-depth and diverse research of higher quality (Fox & Boris, 2016) or use of different expertise from different institutes and countries. However, the association of collaboration with citation counts varies across fields and years. For instance, number of (co)authors associates with a high citation impact in Arts and humanities, Chemistry, Pharmacology, Toxicology, Pharmaceutics (Thelwall & Sud, 2015), and Ecology (Fox & Boris, 2016). In some fields, such as Biomedicine, a positive association of the number of authors with citation impact exists until two or three years after publication and thereafter decreases (Bornmann & Leydesdorff, 2015). Country and level of collaboration (institutional, national, or international) could also affect citation or readership counts. Papers may get more citations due to their larger audience in a given country, or because of their different number of collaborators from different countries (Thelwall & Sud, 2016). These factors could increase visibility of research across different nations and thus associate with higher citation or readership counts. Results of cross-country analysis of Mendeley readers and authors of publications showed that Mendeley users mostly tend to select papers from their own countries, because they are more familiar with authors or fields from their own countries (Thelwall & Maflahi, 2015). High-impact authors, institutes, and countries involved in papers, or high-impact journals in which papers are published could also attract more citations. For instance, Journal Impact Factor (JIF) and international teamwork associate with increased citation impact in certain fields, including Biology and Biochemistry, Chemistry, and Social sciences (Didegah & Thelwall, 2013). However, international collaboration does not necessarily associate with high citation or readership counts. For instance, although the number of institutes and international collaborations associate with decreased citation impact in the field of Biochemistry, only collaboration with high-impact countries, such as the U.S., associates with increased citation and readership counts in this field (Sud & Thelwall, 2016).

In most fields, with the exception of Biomedical and health sciences, a negative relationship between number of pages and citations and Mendeley readership counts exists. This is in contrast, however, to other studies that showed that papers with more pages tend to attract more citations (Haustein et al., 2015; Ajiferuke & Famoye, 2015; Davis, et al., 2011). This contradiction could be due to different sets of publications or choice of statistical tests (simultaneous vs. separate assessment of variables using regression or correlation analyses[8]) used in these studies. Longer papers may present more results and include figures, tables, references, etc., which may indicate a more in-depth analysis or more analytical research. It seems that this is an important feature for authors and Mendeley users in Biomedical and health sciences while in all other fields shorter papers are more attracted to them. It may be the case that the preference for condensed research or easy to read papers are prevalent by both authors and Mendeley users in these fields.

---

[8] In the regression analysis the effect of one variable to another is assessed while controlling for the effect of other variables in the model (simultaneous assessment of variables). In contrast, only the relation between one variable with another variable are assessed by the correlation analysis regardless of controlling for the effect of other variables in the model (separate assessment of variables).

The relationship between title length with citations and readership varies across fields. For instance, in Biomedical and health sciences, Life and earth sciences, and Natural sciences and engineering, readership decreases with longer title length, while citations increase. This may reflect that, opposite to citations, technical titles are not attracted by Mendeley readers. These results are also supported by (Didegah et al., in press) who also found that title length associates with decreased Mendeley reader counts. Similarly, the probability of papers to be tweeted decreases with the lengths of titles (Haustein et al., 2015). Another study found that papers with more characters in their titles received more Mendeley readers than tweets compared to those with fewer characters (Xu, Khalili, & Deng, 2017). However, in Social sciences and humanities longer titles associate with more citation and readership counts as papers with longer titles may inform more specific, detailed, or scientific information in this field. Previous studies have shown that preference for this characteristic varies across different fields, and hence mixed results have been reported. For instance, in fields such as Psychology (Subotic & Mukherjee, 2014; Haslam, et al., 2008) and Marketing (Stremersch, et al., 2015), longer titles associate with fewer citations, while in Medical fields they associate with more citations (Jacques & Sebire, 2010). In PLOS journals, papers with longer titles receive fewer citations and are downloaded less frequently than those with longer titles. However, these results vary for papers with different title characteristics (having colons, question marks, etc.) (Jamali & Nikzad, 2011).

In the results presented above, only a few differences were found between citation and readership counts regarding the relation of number of institutes and title lengths in certain fields, while the results for all other variables were identical to citation impact. We therefore conclude that factors influencing readership and citation counts are broadly similar across all fields. It seems that though differences exist, citation and readership counts are affected by similar variables regardless of whether authors or Mendeley users have the same or different motivations to cite or save documents. This is to be expected due to the similarities between Mendeley users (mostly academics) and authors. From a more conceptual point of view, the broadly similar relationship between readership and citations can be explained by the use of Mendeley in a pre-citation context (Haustein, Bowman, & Costas, 2016) and by its stronger use among academic users (Zahedi, Costas, & Wouters, 2014). A survey of Mendeley users showed that this tool is used mostly by students, and that most Mendeley users use the tool to cite literature in their publications, followed by professional purposes (i.e. updating for job), teaching, and educational activities (Mohammadi, Thelwall, & Kousha, 2016).

However, both citation and readership counts are influenced by a range of motivations depending on the preference of authors and users when citing and saving documents. The act of saving documents in Mendeley can be also linked to motivations different from those for citing[9], for example reading for self-awareness, teaching, curiosity, or other professional needs (Haustein et al., 2015). This explains for example why some document types are substantially saved in Mendeley although they are not particularly cited later on (e.g., editorial materials, letters, book reviews, meeting abstracts, or news items). Therefore, it can be argued that although citing and saving are related activities, they also have some fundamental differences. Hence, Mendeley users do not always "adhere to the same norms as citations" when saving a document (Haustein et al., 2015). Moreover, both citation and readership impact could be influenced by some random (i.e. perfunctory) effects (Waltman, van Eck, & Wouters, 2013) that contribute to reducing the reliability of both readership and citation counts. Overall, the existence of these different motivations for saving documents in Mendeley and citing them suggests that one should not fully associate readership with citation counts. However, understanding all possible reasons for an author or Mendeley user to prefer some documents to others when citing and/or saving documents requires a more qualitative analysis; which is beyond the scope of the present study. Moreover, all the above results could be influenced by properties related to other variables (for instance author's gender, country, and reputation or impact; cited reference's impact and recency, etc. as discussed by Stremersch, et al., 2015; Bornmann et al., 2012) and by specific norms and peculiarities of each field. Furthermore, the relationships between documents characteristics and citation and readership counts found in this study don't imply causation.

---

[9] For a review on citation behaviour see Bornmann & Daniel (2008).

Finally, the limitations of this study need to be acknowledged. One limitation is that the variables analyzed are restricted to the same variables as studied by Haustein et al. (2015) in order to have comparable values to other altmetric sources. However, there could be other cultural, technological, economical, and political variables that could have an important effect on citation and readership counts. The extent to which researchers decide to use online platforms (like Mendeley) in their research workflow is influenced by many other external factors (Jamali, et al., 2015; Gruzd, Staves, Wilk, 2012; Rowlands et al., 2011), such as different levels of familiarity with online platforms, or the age of researchers (e.g., younger researchers use social media more often than older researchers (Bolton et al., 2013; Bowman, 2015; Tenopir et al., 2015; Nicholas et al., 2015). For example, the different use and adoption of social media by different communities (as discussed by Bolton et al., 2013) could play an important role in why for some areas and some document types have a stronger (or weaker) relationship with Mendeley readership. Moreover, other document characteristics- such as openness (e.g. being open access), language of the publication, database coverage, availability of DOIs, or other different identifiers for the same document, the lack of complete metadata, etc.- may directly influence the rate of citation and readership received by a paper (Thelwall, 2015; Zahedi, Bowman, & Haustein, 2014). From a research evaluation perspective, one also needs to consider the underlying biases of Mendeley. Hence, there are age biases as Mendeley is used by younger researchers (Haustein & Larivière, 2014). There are also geographical biases in the coverage of publications, with some countries strongly covered and represented on Mendeley (Thelwall & Maflahi, 2015). All these biases call for caution in the interpretation of factors that could influence readership counts, particularly regarding the practical application of this type of indicator in any applied context.

**Acknowledgements**

The authors would like to thank the anonymous reviewers for their comments on this paper. Special thanks to Timothy Bowman from Wayne State University for his comments on the paper. The idea for this study came out when Zohreh Zahedi was a visiting scholar at the Canada Research Chair on the Transformations of Scholarly Communication (Université de Montréal), during Summer 2014, supported partly by the grant awarded by the Leiden University Fund (LUF #4509) and partly by funding from the Iranian Ministry of Science, Research, & Technology scholarship program (MSRT grant #89100156).

## Appendices

**Table A1.** Descriptive table of the independent and the dependent variables across articles and reviews disciplines.

| Leiden Ranking Discipline | | PG | NR | TI | AU | IN | CU |
|---|---|---|---|---|---|---|---|
| **All disciplines** | Density | 10.06 | 39.60 | 98.75 | 5.42 | 2.21 | 1.34 |
| N=1,197,162 | *Std. dev.* | 7.60 | 31.22 | 35.05 | 37.70 | 3.75 | 0.98 |
| **Biomedical & health sciences** | Density | 8.75 | 41.30 | 104.51 | 5.98 | 2.45 | 1.33 |
| N=502,610 | *Std. dev.* | *6.81* | *33.03* | *36.51* | *5.11* | *2.50* | *0.88* |
| **Life & earth sciences** | Density | 10.28 | 45.55 | 106.43 | 4.96 | 2.28 | 1.41 |
| N=242,117 | *Std. dev.* | *6.56* | *31.20* | *33.88* | *4.27* | *1.78* | *0.83* |
| **Mathematics & computer science** | Density | 13.06 | 28.14 | 81.91 | 3.21 | 1.82 | 1.31 |
| N=131,220 | *Std. dev.* | *8.40* | *19.32* | *27.64* | *2.81* | *1.16* | *0.61* |
| **Natural sciences & engineering** | Density | 8.94 | 36.08 | 97.53 | 6.38 | 2.12 | 1.35 |
| N=404,457 | *Std. dev.* | *6.58* | *29.13* | *33.06* | *64.55* | *5.70* | *1.25* |
| **Social sciences & humanities** | Density | 15.02 | 46.32 | 87.15 | 2.77 | 1.83 | 1.26 |
| N=125,792 | *Std. dev.* | *8.18* | *30.16* | *32.10* | *2.45* | *1.38* | *0.64* |

PG= Number of Page; NR= Number of Reference, TI=Title Length, AU=Number of Author, IN=Number of Institute, CU=Number of Country.

**Table A2.** Spearmann correlation analysis among the independent and the dependent variables (C and MR) across articles and reviews by disciplines.

**Biomedical & health sciences (n= 502,610)**

| | PG | NR | TI | AU | IN | CU | C | MR |
|---|---|---|---|---|---|---|---|---|
| **PG** | 1.000 | 0.652 | 0.092 | 0.052 | 0.105 | 0.132 | 0.318 | 0.338 |
| **NR** | | 1.000 | -0.008 | -0.031 | 0.064 | 0.115 | 0.381 | 0.392 |
| **TI** | | | 1.000 | 0.253 | 0.107 | 0.046 | 0.052 | -0.033 |
| **AU** | | | | 1.000 | 0.452 | 0.243 | 0.186 | 0.015 |
| **IN** | | | | | 1.000 | 0.511 | 0.156 | 0.122 |
| **CU** | | | | | | 1.000 | 0.165 | 0.164 |
| **C** | | | | | | | 1.000 | 0.558 |
| **MR** | | | | | | | | 1.000 |

**Life & earth sciences (n=242,117)**

| | PG | NR | TI | AU | IN | CU | C | MR |
|---|---|---|---|---|---|---|---|---|
| **PG** | 1.000 | .560 | .069 | -.041 | .076 | .106 | .149 | .166 |
| **NR** | | 1.000 | .067 | .031 | .107 | .133 | .345 | .383 |
| **TI** | | | 1.000 | .156 | .064 | .019 | .008 | -.065 |
| **AU** | | | | 1.000 | .508 | .272 | .196 | .070 |
| **IN** | | | | | 1.000 | .579 | .142 | .132 |

**Natural sciences & engineering (n=404,457)**

| | PG | NR | TI | AU | IN | CU | C | MR |
|---|---|---|---|---|---|---|---|---|
| **PG** | 1.000 | .424 | .020 | -.127 | .051 | .087 | .081 | .095 |
| **NR** | | 1.000 | .101 | .061 | .089 | .121 | .429 | .341 |
| **TI** | | | 1.000 | .157 | .006 | -.043 | .073 | 0.001 |
| **AU** | | | | 1.000 | .434 | .246 | .185 | .132 |
| **IN** | | | | | 1.000 | .597 | .097 | .087 |
| **CU** | | | | | | 1.000 | .126 | .110 |
| **C** | | | | | | | 1.000 | .581 |
| **MR** | | | | | | | | 1.000 |

**Mathematics & computer sciences (n= 131,220)**

| | PG | NR | TI | AU | IN | CU | C | MR |
|---|---|---|---|---|---|---|---|---|
| **PG** | 1.000 | .430 | -.071 | -.168 | .054 | .095 | .022 | .034 |
| **NR** | | 1.000 | .122 | .122 | .136 | .111 | .336 | .424 |
| **TI** | | | 1.000 | .184 | .040 | -.027 | .119 | .080 |
| **AU** | | | | 1.000 | .454 | .251 | .193 | .242 |
| **IN** | | | | | 1.000 | .601 | .119 | .123 |
| **CU** | | | | | | 1.000 | .104 | .091 |
| **C** | | | | | | | 1.000 | .452 |
| **MR** | | | | | | | | 1.000 |

**Social sciences & humanities (n=125,795)**

| | PG | NR | TI | AU | IN | CU | C | MR |
|---|---|---|---|---|---|---|---|---|
| **PG** | 1.000 | .412 | -.041 | -.264 | -.114 | -.017 | -.103 | -.081 |
| **NR** | | 1.000 | .161[**] | .075 | .079 | .073 | .286 | .322 |
| **TI** | | | 1.000 | .273 | .146 | .043 | .157 | .138 |
| **AU** | | | | 1.000 | .635 | .326 | .396 | .339 |
| **IN** | | | | | 1.000 | .570 | .289 | .247 |
| **CU** | | | | | | 1.000 | .178 | .175 |
| **C** | | | | | | | 1.000 | .638 |
| **MR** | | | | | | | | 1.000 |

PG= Number of Page; NR= Number of Reference, TI=Title Length, AU=Number of Author, IN=Number of Institute, CU=Number of Country, C= Citation, MR= Mendeley Readership